\documentclass[epj,nopacs]{svjour}
\pdfoutput=1
\usepackage{graphics}
\usepackage {placeins}
\usepackage{amsmath}
\usepackage{amssymb}
\usepackage{subfigure}
\usepackage{lineno}
\usepackage{hyperref}
\begin{document}
%
\title{Spatially Resolved Dark Count Rate of SiPMs}
\author{E. Engelmann\inst{1}\thanks{\emph{Present address:} eug.engelmann@gmx.de}
\and E. Popova \inst{2}
\and S. Vinogradov \inst{2,}\inst{3}
%
}                     
\offprints{}          
\institute{Institute of Physics, Universität der Bundeswehr München, Munich, Germany
\and National Research Nuclear University MEPhI, Moscow, Russia
\and P. N. Lebedev Physical Institute, Moscow, Russia}
\date{Received: date / Revised version: date}
%
\abstract{
The Silicon Photomultiplier (SiPM) is a promising photo-detector for a variety of applications.
However, the high dark count rate ($DCR$) of the SiPM is still a contemporary problem. Decreasing the $DCR$ would significantly broaden the range of possible applications.
In this work we present a novel method for the spatially resolved characterization of crystal defects in SiPMs.
The contribution of crystal defects to the $DCR$ is evaluated by exploiting the effect of "hot carrier luminescence" (HCL), which is light that is emitted during the Geiger mode operation of avalanche photodiodes (SiPM micro-cells).
Spatially confined regions with an enhanced light emission intensity (hotspots) are identified within the active areas of SiPM micro-cells. By correlating the detected light intensity and the $DCR$, a significant contribution of up to $56\text{ \%}$ of the $DCR$ can be attributed to less than $5\text{ \%}$ of the micro-cells.
The analysis of the temperature dependence of the emitted light identifies the Shockley-Read-Hall-Generation to be the dominant mechanism responsible for the occurrence of hotspots.
The motivation of this work is to generate a deeper understanding of the origin of hotspots in order to suppress their contribution to the $DCR$ of SiPMs. 
%
} 
\maketitle
\section{Introduction}
\label{sec_Introduction}
The emission of light during avalanche breakdowns in reverse biased p-n junctions was observed and investigated by various authors. D. K. Gautam et al. \cite{Gautam} modeled the light emission spectrum by indirect recombination of electrons and holes under ionizing conditions. For avalanche photodiodes, A. L. Lacaita et al. reported an intensity of $2.9 \cdot 10^{-5}$ photons with wavelength lower than $1088\text{ nm}$ per carrier crossing the junction \cite{Lacaita1993}. The process responsible for the hot carrier luminescence (HCL) was attributed to electron (hole) energy relaxations between states of the conduction (valence) band. For SiPMs, R. Mirzoyan et al. \cite{Mirzoyan} reported an intensity of $2.6\cdot 10^{-5}$ photons per electron in the spectral range from $500\text{ nm}$ to $1600\text{ nm}$ for Hamamatsu MPPC \textit{S10362-11-10U}.
Mapping of the dark count rate ($DCR$) for every micro-cell of a digital SiPM was reported by T. Frach et al. \cite{dSiPM}. The group showed that excluding $5\text{\%}$ of the most active micro-cells results in a reduction of the dark count rate by up to an order of magnitude. Their results did not provide information on the origin of micro-cells with a higher $DCR$.
In this work we present a method for a spatially resolved characterization of the $DCR$ of SiPMs by detecting light emitted by HCL with a low-light level CCD camera. The applicability of this method is demonstrated for two KETEK SiPM types \cite{KETEK}.
\FloatBarrier
\begin{figure}
\centering
\resizebox{0.45\textwidth}{!}{%
  \includegraphics{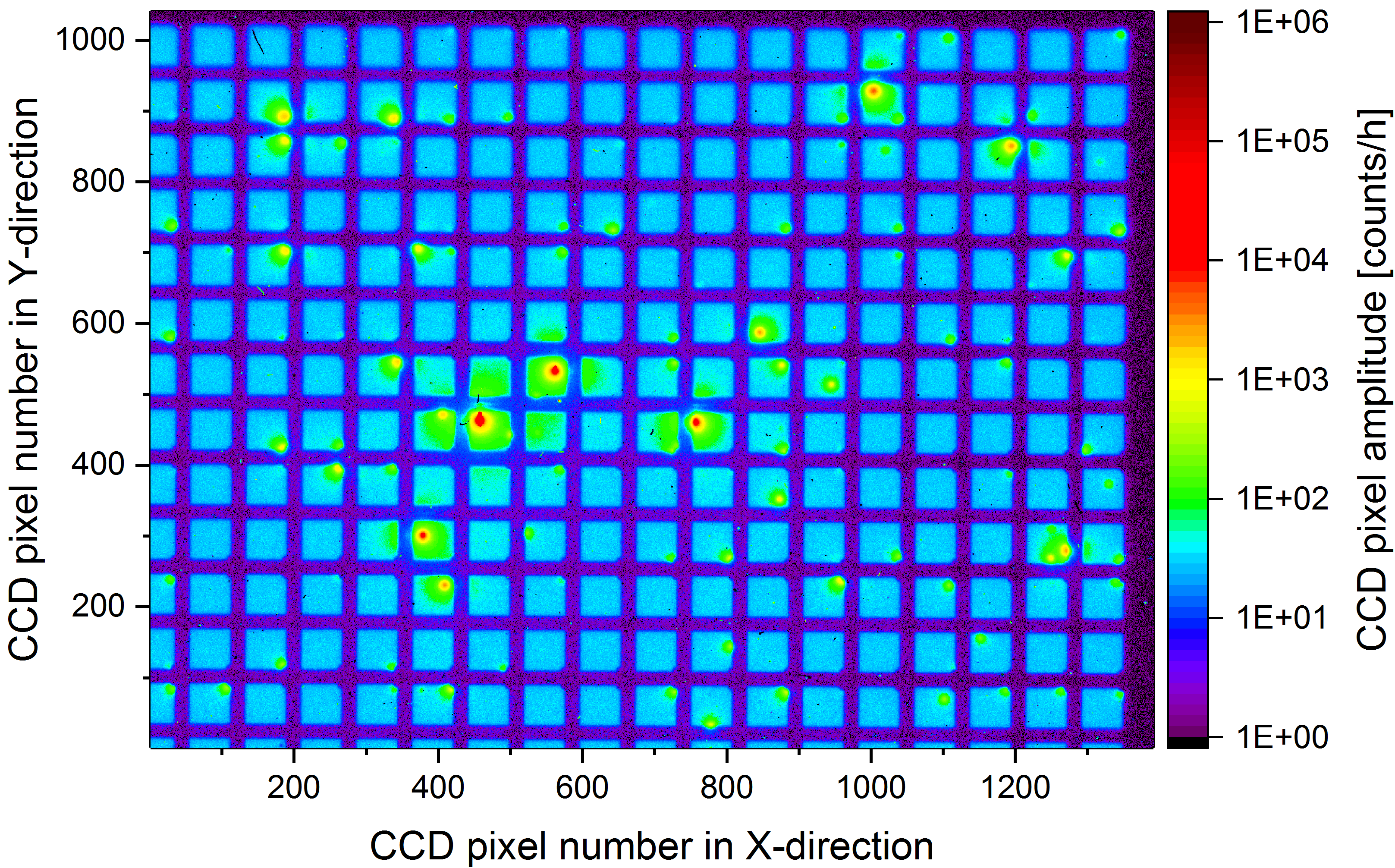}
}
\caption{Light emission image of the PM3350T STD at $t_{exp}=4\text{ h}$ and $\Delta V=5.4\text{ V}$. Active areas of single micro-cells (blue) and hotspots within the micro-cells (red) are clearly visible.}
\label{CCD_examples_b}       
\end{figure}
\FloatBarrier
Figure \ref{CCD_examples_b} shows a light emission image of a PM3350T STD at an excess bias voltage of $\Delta V=V-V_{bd}=5.4\text{ V}$. Here, $V$ is the reverse bias voltage and $V_{bd}$ is the breakdown voltage of the SiPM. The active areas of micro-cells can be clearly distinguished from the non-emitting metal lines.
Furthermore, regions with a significantly enhanced light intensity (hotspots) are evident within the active areas of certain SiPM micro-cells. 
Figure \ref{fig_Intensity_Map} shows the relative contribution of each micro-cell to the overall detected light intensity in the field of view.
Micro-cells with a hotspot contribute with up to $7\text{ \%}$ to the total $DCR$, while a contribution of approximately $0.45\text{ \%}$ is expected on average.
The goal of our work is to correlate the spatially resolved light intensity with the $DCR$ and to determine the contribution of hotspot-generated dark count rate to the total $DCR$.
\begin{figure}
\centering
\resizebox{0.5\textwidth}{!}{%
  \includegraphics{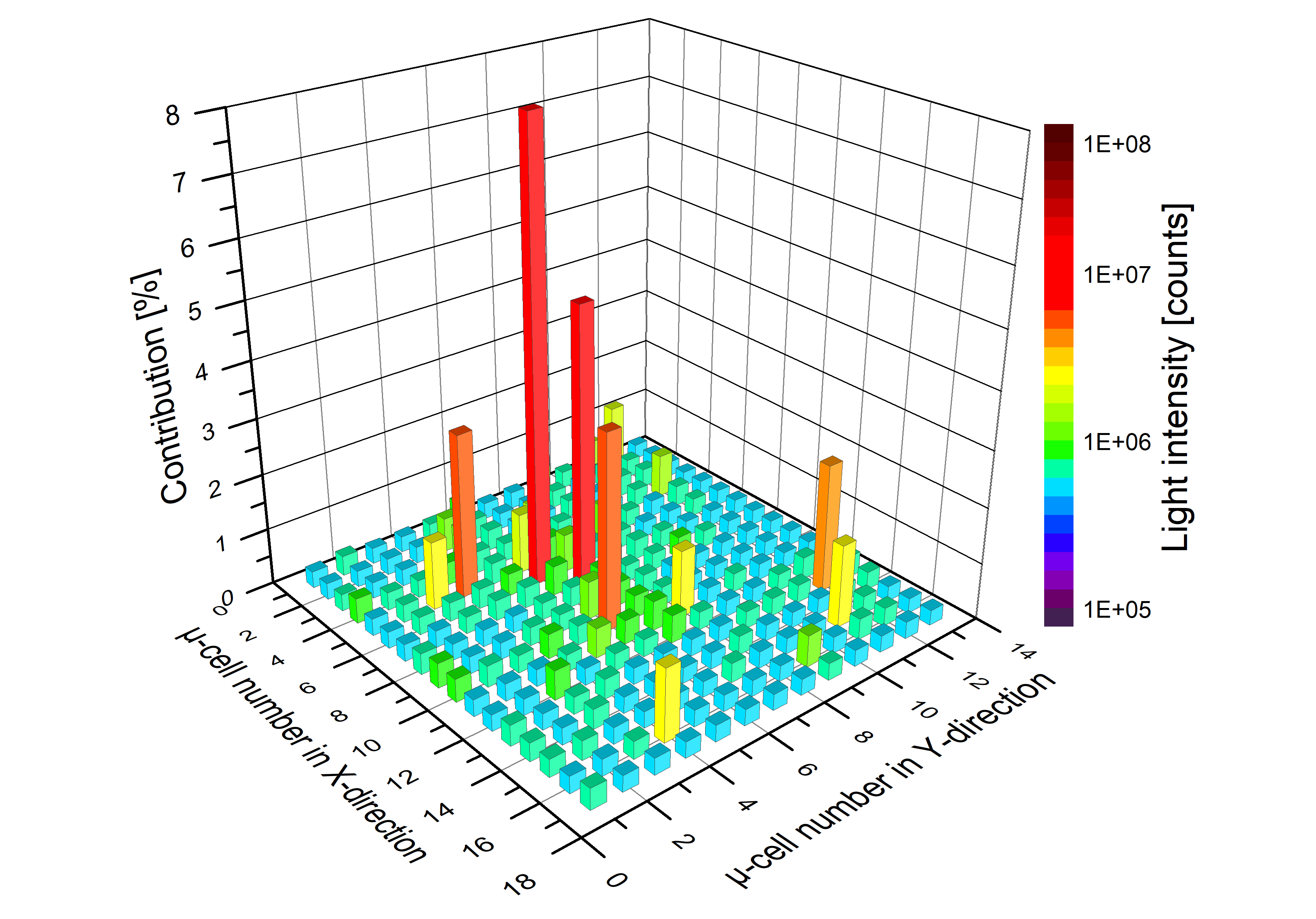}
}
\caption{Contribution of each micro-cell from figure \ref{CCD_examples_b} to the overall emitted light intensity.}
\label{fig_Intensity_Map}       
\end{figure}
\FloatBarrier
\section{Experimental setup}
\label{sec_Experimental_setup}
\begin{figure}
\centering
\resizebox{0.45\textwidth}{!}{%
  \includegraphics{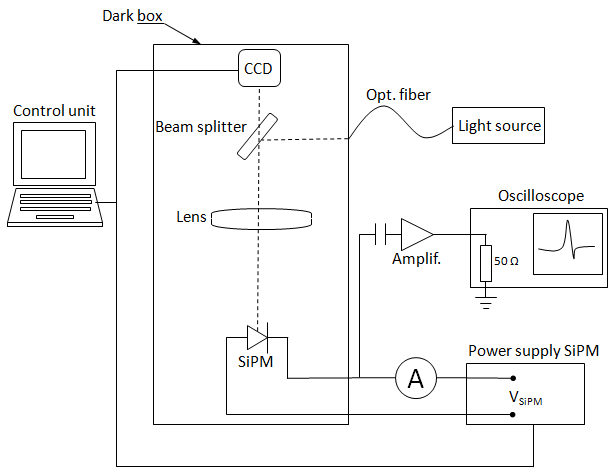}
}
\caption{Sketch of the experimental setup.}
\label{Setup_Schematics}       
\end{figure}

For the measurement the SiPM is placed inside a dark box.
The light emitted during avalanche breakdowns of micro-cells is detected with the low-light level CCD camera "\textit{Clara}" with enhanced near-infrared sensitivity, produced by \textit{Andor} \cite{Andor}. 
The camera is attached to an optical microscope \textit{Mitutoyo FS70 S/N}. A sketch of the setup is shown in figure \ref{Setup_Schematics}.
For measurements at room temperature, the exposure time $t_{exp}$ is set to 1 h, which is sufficiently long to discriminate between light emitted from the SiPM and thermal noise of the CCD. A longer exposure time can lead to a saturation of the CCD due to the light emitted by intense hotspots. For lower temperatures, $t_{exp}$ is increased up to 5 h due to the fast decrease of the SiPM dark count rate.
Every image is background corrected by subtracting a CCD image which is recorded with the SiPM bias voltage set to zero. The presented measurements were performed in a temperature stabilized environment at $(21\pm1)\text{ }^{\circ}\text{C}$, unless otherwise stated.
\FloatBarrier
\section{Investigated samples}
\label{sec_Investigated samples}
\begin{figure}[h!]
\centering
\resizebox{0.5\textwidth}{!}{%
  \includegraphics{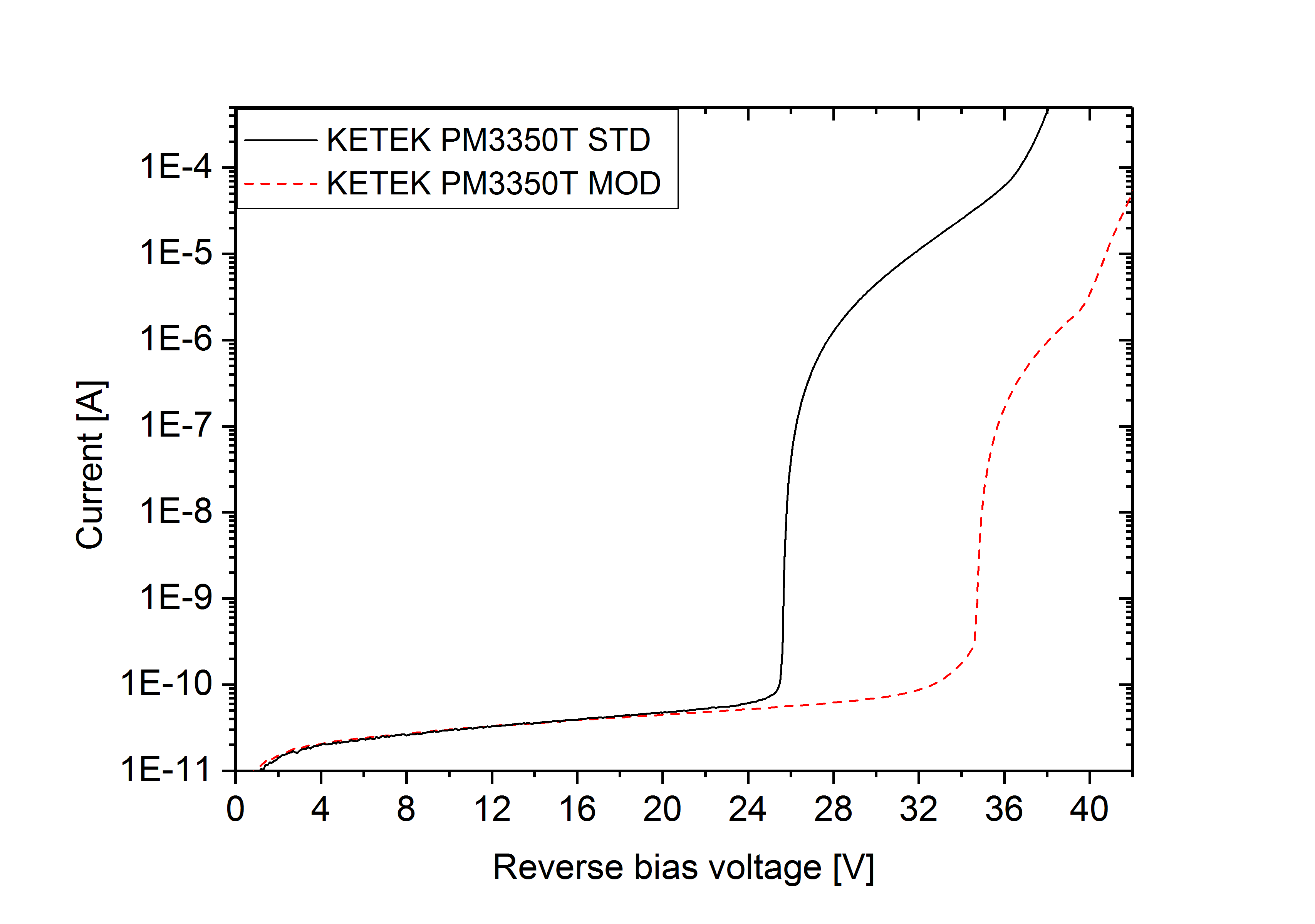}
}
\caption{Current-voltage characteristics of the KETEK PM3350T STD and the PM3350T MOD at dark conditions.}
\label{fig_IV_STD_HE}       
\end{figure}

For this study we use our method to characterize two blue sensitive SiPM types from KETEK. The parameters of the investigated samples are listed in table \ref{tab_samples_list_ImplEnergy}. The key difference between both devices is the increased breakdown voltage for the PM3350T MOD due to a thicker p-n junction.
Figure \ref{fig_IV_STD_HE} shows examples of the current-voltage characteristics of both SiPMs at dark conditions.
\begin{table}[h!]
\caption{List of investigated samples}
\label{tab_samples_list_ImplEnergy}       
\begin{tabular}{lll}
\hline\noalign{\smallskip}
SiPM & PM3350T STD & PM3350T MOD\\
\noalign{\smallskip}\hline\noalign{\smallskip}
Breakdown voltage $V_{bd}$ & $(25.6\pm0.1)\text{ V}$& $(34.6\pm0.1)\text{ V}$\\
Number of micro-cells & $3600$& $3600$\\
Active area & $9\text{ mm}^2$& $9\text{ mm}^2$\\
Micro-cell pitch & $50\text{ }\mu \text{m}$& $50\text{ }\mu \text{m}$\\
Recovery time & $(80.8\pm0.1)\text{ ns}$& $(65.8\pm0.1)\text{ ns}$\\ 
Number of samples & 6& 6\\  
\noalign{\smallskip}\hline
\end{tabular}
\end{table}
\FloatBarrier
In the following, the photon detection efficiency ($PDE$) at $406\text{ nm}$ of the investigated devices is compared.
For this measurement we use the statistical analysis reported in \cite{Eckert2010}, \cite{Otte2016}.
In the first step, the SiPM is illuminated with a fixed light intensity from a pulsed laser (\textit{PLP-10, Hamamatsu}, $406\text{ nm}$, $<70\text{ ps}$ pulse width).
A total number of $N_{tot}$ laser flashes are recorded with an oscilloscope. Assuming that the number of photons in each light pulse follows a Poisson distribution, the average number of detected photons per light pulse is given by $-ln(N_0/N_{tot})$. Here, $N_{0}$ is the number of light pulses for which no photon was detected.
In the second step, the measurement is repeated without actually flashing the SiPM. The average number of detected dark pulses is given by $-ln(N_{0}^{dark}/N_{tot})$, with $N_{0}^{dark}$ being the number of recorded traces without the SiPM detecting a dark pulse.
The resulting dark-count corrected average number of detected photons per light pulse $\mu$ is given in equation \ref{eq_PDE}. 
\begin{equation}
\label{eq_PDE}
PDE\sim\mu=-ln\left(\frac{N_{0}}{N_{tot}}\right)+ln\left(\frac{N_{0}^{dark}}{N_{tot}^{dark}}\right)
\end{equation}
\begin{figure}
\centering
\resizebox{0.5\textwidth}{!}{%
  \includegraphics{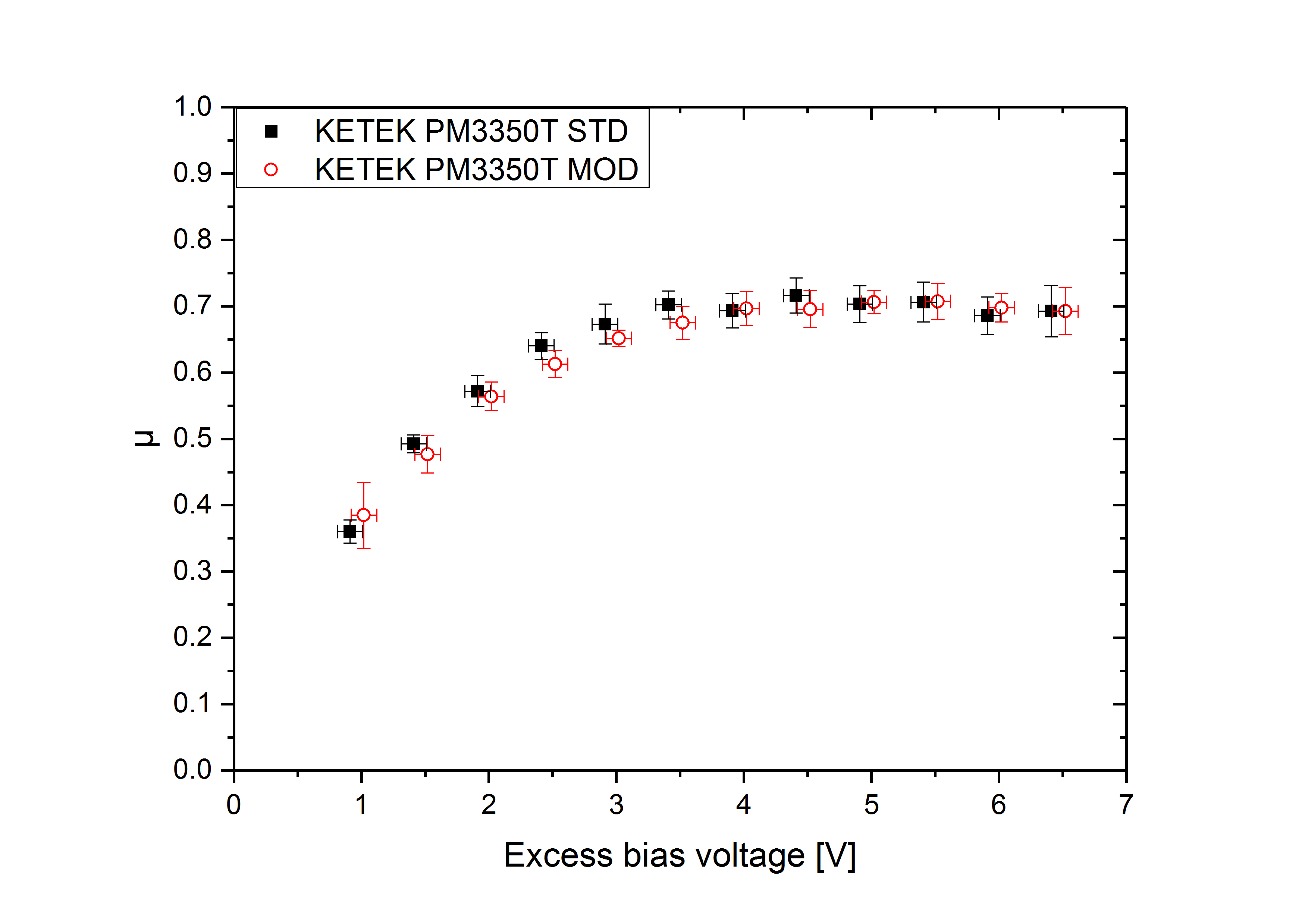}
}
\caption{Average number of detected photons per light pulse $\mu$.}
\label{fig_PDE_STD_MOD}       
\end{figure}
Both SiPMs show a similar dependence of $\mu$ on the excess bias voltage within the uncertainties. Due to an equal geometrical efficiency we conclude a similar $PDE(\Delta V)$ for both SiPMs.
For this reason the SiPMs are compared at a fixed $\Delta V$.
\FloatBarrier
\section{Algorithm for CCD-data analysis}
\label{sec_Algorithm_for_CCD-data_analysis}
In this section the developed characterization method is described on the example of the PM3350T STD. The number of hotspots and the light intensity emitted from each hotspot is determined. The emitted light intensity is correlated with the $DCR$. Based on this correlation, the contribution of the $DCR$ generated in hotspots to the total dark count rate of the SiPM is determined.
\subsection{Determination of $I_{glowing}$ and $I_{hotspots}$}
\label{sub_Iglowing_Ihotspots}
\begin{figure}
\centering
\resizebox{0.5\textwidth}{!}{%
  \includegraphics{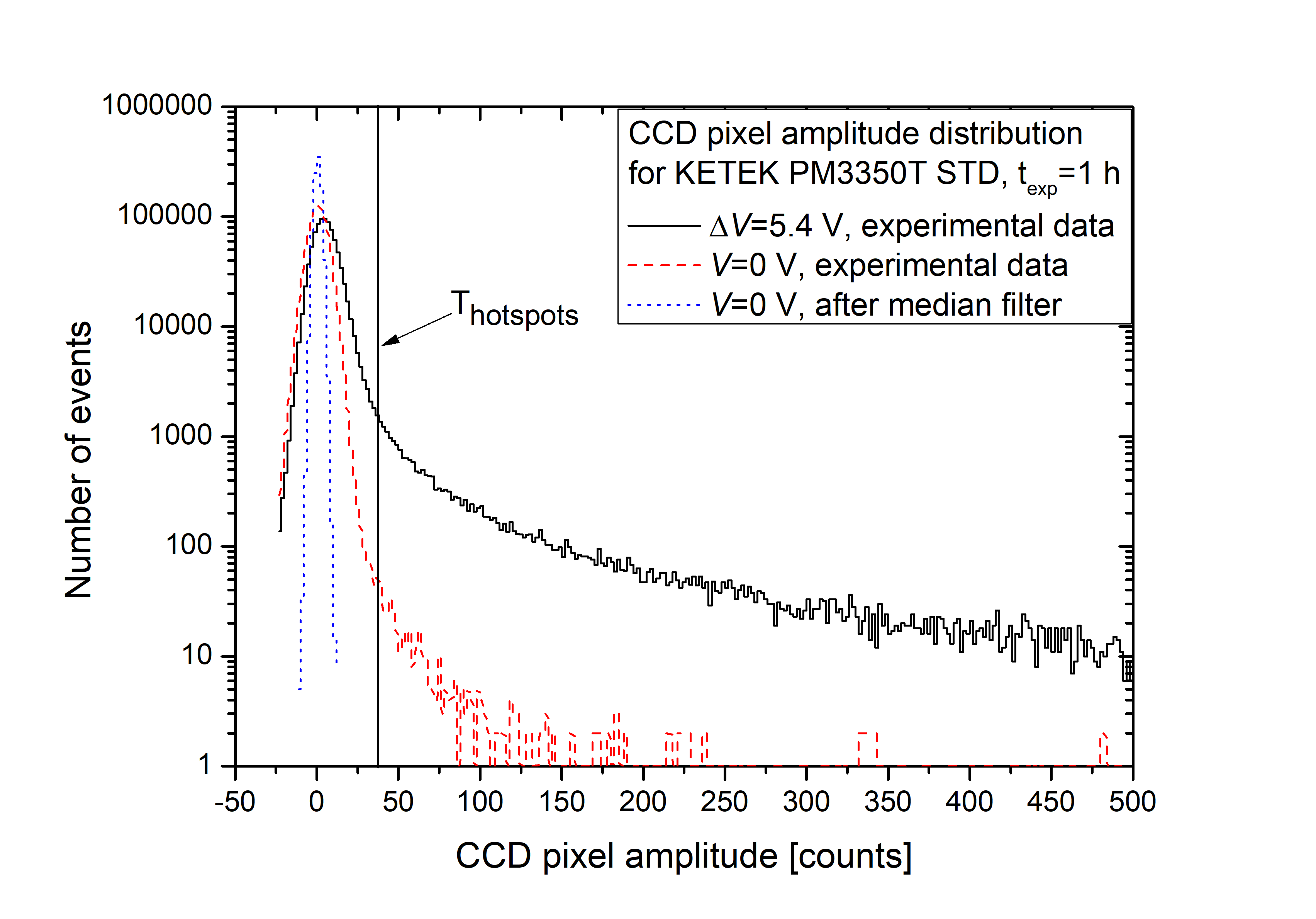}
}
\caption{Examples of CCD pixel amplitude distributions.}
\label{fig_Spectra_Example_biased_unbiased}       
\end{figure}

\begin{figure} [h!]
\centering
\resizebox{0.5\textwidth}{!}{%
  \includegraphics{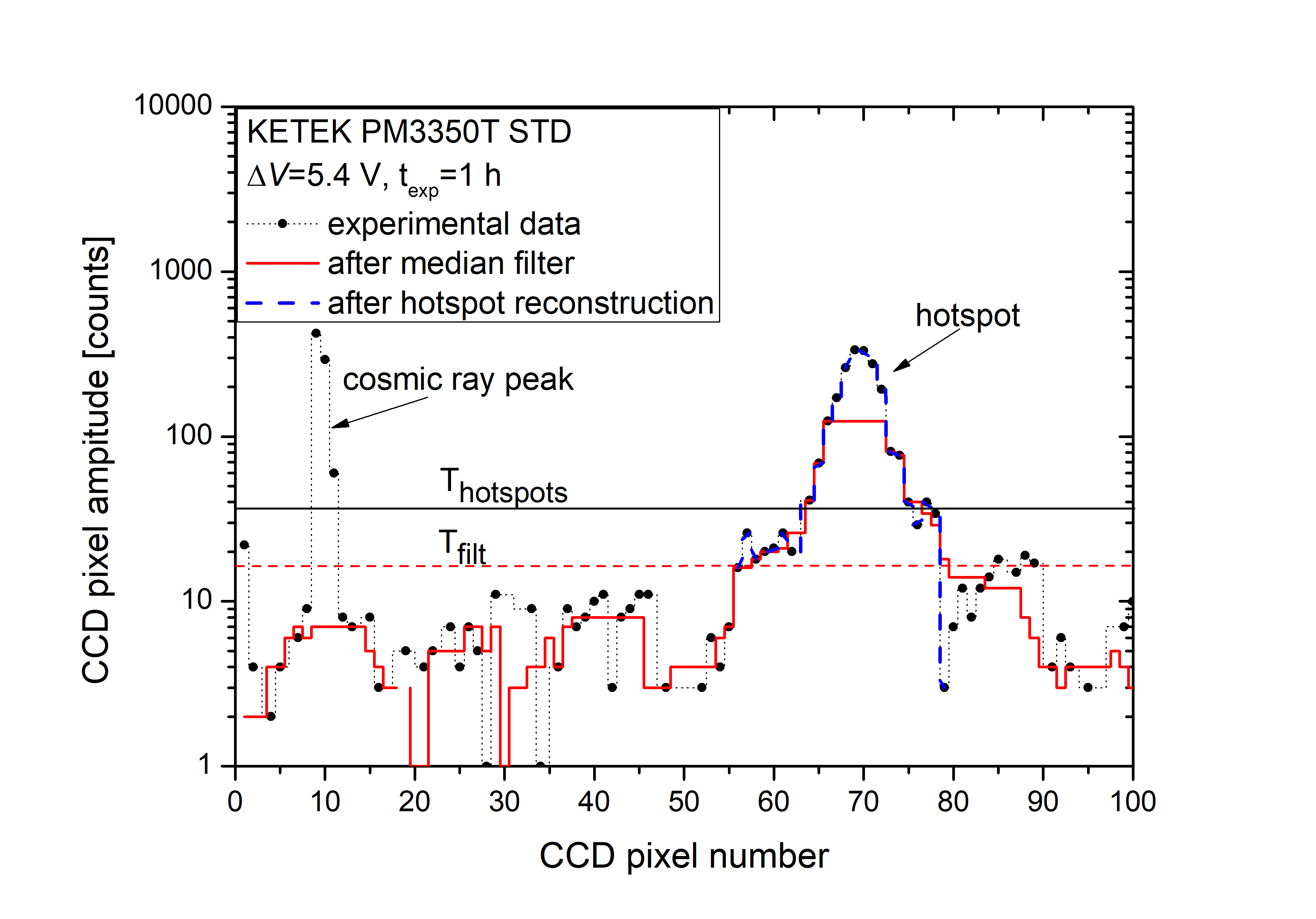}
}
\caption{Example of a hotspot and a cosmic ray peak.}
\label{CCD_different_stages}       
\end{figure}

\begin{figure}[h!]
\centering
\resizebox{0.5\textwidth}{!}{%
  \includegraphics{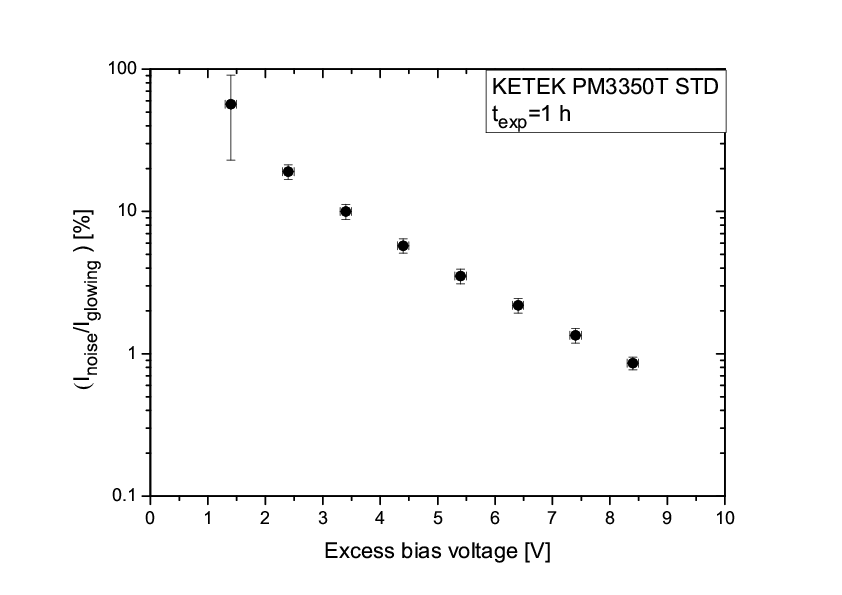}
}
\caption{Contribution of the CCD noise $I_{noise}$ to $I_{glowing}$.}
\label{fig_Contribution_CCDnois_to_Iglowing}       
\end{figure}

The camera signal is divided into two main contributions. The first contribution is attributed to the homogeneous emission of light from every micro-cell without hotspots (blue regions in figure \ref{CCD_examples_b}). We call this contribution "glowing intensity" $I_{glowing}$ and define it as the sum of all CCD pixel amplitudes below a hotspot threshold $T_{hotspots}$. The second contribution is due to hotspots and is determined as the sum of all pixel amplitudes above $T_{hotspots}$. We call this quantity "hotspots intensity" $I_{hotspots}$. Equations \ref{eq_I_glowing} and \ref{eq_I_hotspots} give a formal definition of both quantities, where $A_{i,j}$ is the experimentally determined CCD pixel amplitude at position (i,j).
\begin{equation}
\label{eq_I_glowing}
I_{glowing}=\sum_{i=1}^n\sum_{j=1}^m A_{i,j} \text{ , for } A_{i,j} \leq T_{hotspots}
\end{equation}
\begin{equation}
\label{eq_I_hotspots}
I_{hotspots}=\sum_{i=1}^n \sum_{j=1}^m A_{i,j} \text{ , for } A_{i,j} > T_{hotspots}
\end{equation}
Figure \ref{fig_Spectra_Example_biased_unbiased} shows an example of the CCD pixel amplitude spectrum for the PM3350T STD. The spectrum is continuous and does not provide a feature that allows to define an universal threshold $T_{hotspots}$, which separates pixels with hotspots from those without. For this reason an arbitrary method is introduced to distinguish between $I_{glowing}$ and $I_{hotspots}$. In the presented method we determine $T_{hotspots}$ separately for each SiPM and operating voltage.
The left part of the spectrum is attributed to the sum of $I_{glowing}$ and the CCD noise. This part is fitted with a Gaussian and $T_{hotspots}$ is set at 4 standard deviations from the mean of the distribution. This is indicated by the vertical line in figure \ref{fig_Spectra_Example_biased_unbiased}. 
Figure \ref{CCD_different_stages} shows an example of pixel amplitudes along one CCD row. It is evident from the figure that the chosen threshold provides a clear separation between hotspots and hotspot-free regions.\\
During the data acquisition, cosmic rays interacting in the CCD camera generate additional signals with CCD pixel amplitudes comparable to hotspots. In the left part of figure \ref{CCD_different_stages} an example of a cosmic-ray peak is shown. In figure \ref{fig_Spectra_Example_biased_unbiased} the influence of cosmic rays on the CCD signal is evident as the non-Gaussian tail of the CCD pixel amplitude distribution at zero bias voltage (dashed curve).
Considering the fact that cosmic-ray signals appear as peaks much narrower ($\approx 1-3$ CCD pixels) than hotspots ($>10$ CCD pixels), a 13th-order, one-dimensional median filter from the LabVIEW library is used to exclude these peaks from the analysis.
The dotted curve in figure \ref{fig_Spectra_Example_biased_unbiased} shows the CCD pixel amplitude distribution at zero bias voltage after the application of the median filter.
The suppression of the cosmic-ray peaks is clearly evident. 
However, the median filter also significantly modifies the amplitudes and shapes of hotspots as indicated by the solid line in figure \ref{CCD_different_stages}. 
For this reason, the filtered CCD pixel amplitudes $A_{i,j}^{filt}$ are only used to determine the position and the lateral expansion of hotspots. For this purpose the threshold $T_{filt}$ is set analogous to the method described for $T_{hotspots}$. CCD pixel amplitudes $A_{i,j}^{filt}$ larger than $T_{filt}$ are attributed to hotspots.
In order to determine the hotspot contribution to the total light intensity, the initial CCD pixel amplitudes $A_{i,j}$ within the lateral boundaries of hotspots are used. The dashed line in figure \ref{CCD_different_stages} indicates the part of the CCD signal which is attributed to a hotspot.
\FloatBarrier
Figure \ref{fig_Contribution_CCDnois_to_Iglowing} shows the contribution of the CCD noise $I_{noise}$ to $I_{glowing}$. The contribution decreases with the excess bias voltage and is below $5 \text{ \%}$ at $\Delta V=5.4\text{ V}$.
\subsection{Counting hotspots}
\label{sub_Counting_hotspots}
\begin{figure}[h!]
\centering
\resizebox{0.5\textwidth}{!}{%
  \includegraphics{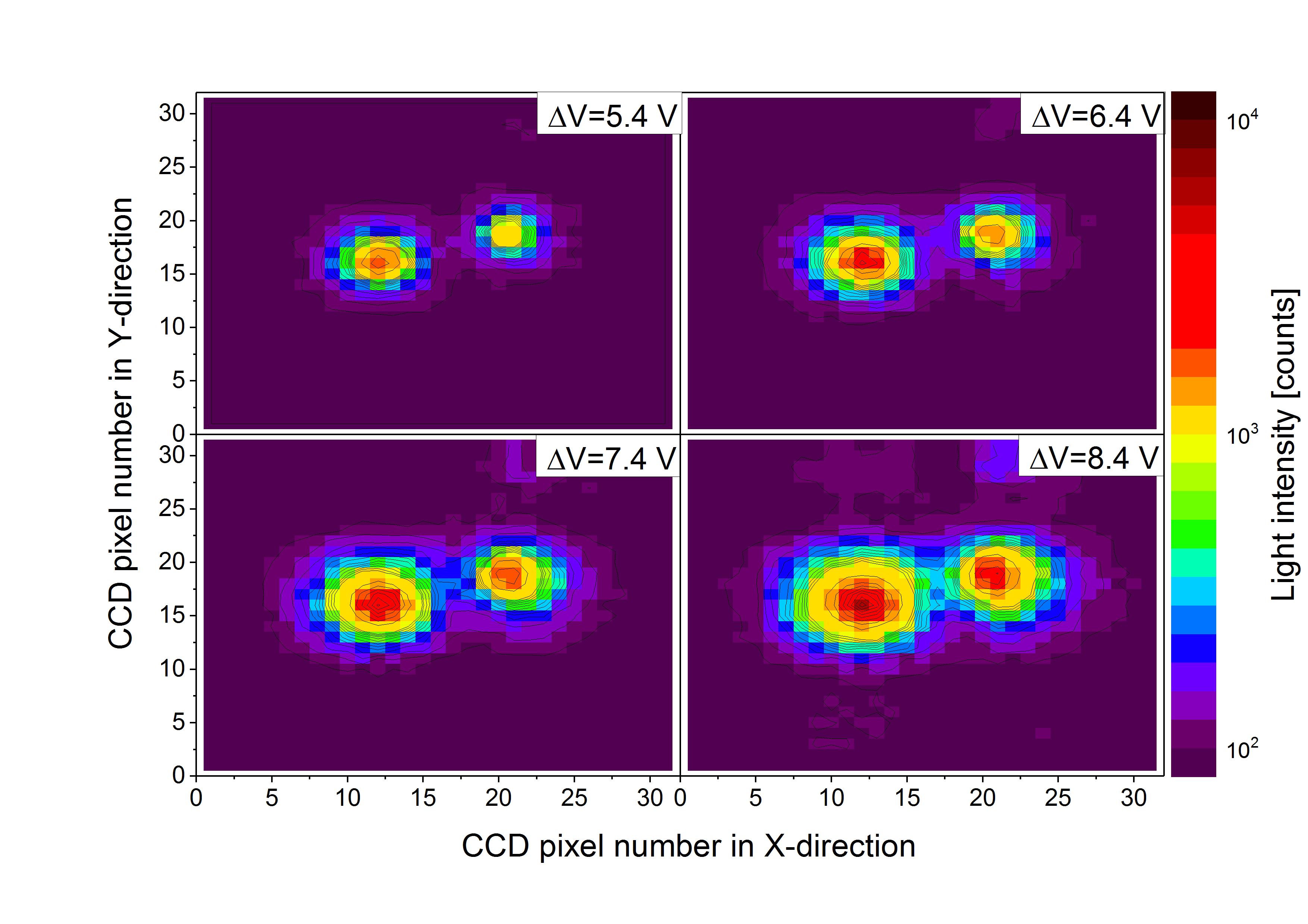}
}
\caption{Example of the increase in lateral expansion of hotspots with $\Delta V$ for a KETEK PM3350T STD ($t_{exp}=1\text{ h}$).}
\label{fig_Ampl_Distr_Two_HS}       
\end{figure}

\begin{figure}[h!]
\centering
\resizebox{0.5\textwidth}{!}{%
  \includegraphics{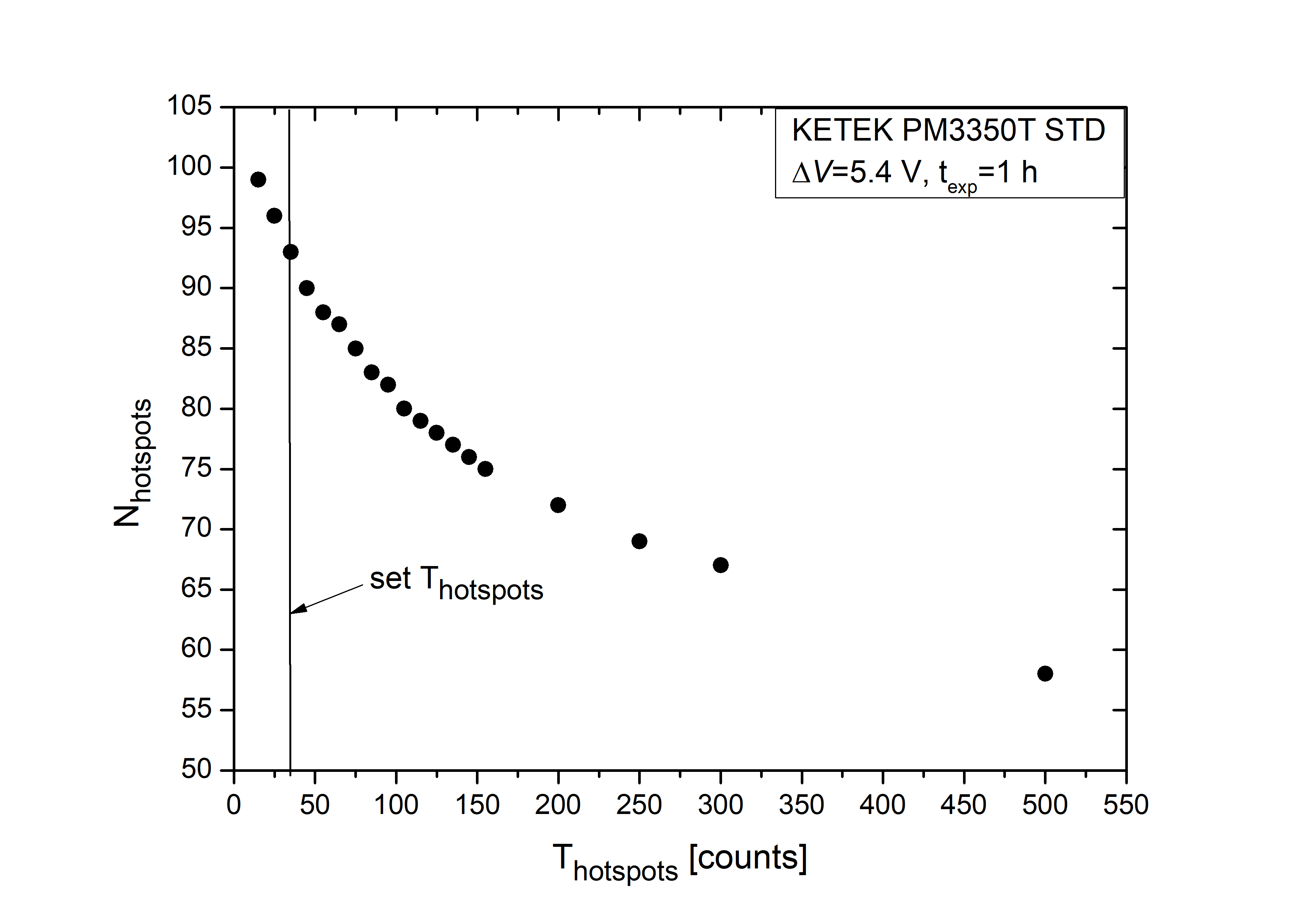}
}
\caption{Dependence of $N_{hotspots}$ on $T_{hotspots}$.}
\label{fig_Number_Hotspots(Th_hotspots)}       
\end{figure}

The number of hotspots and the light emission intensity of each hotspot is determined in three steps.
In the first step, the position of hotspot centers is determined by identifying the coordinates of local amplitude maxima in the CCD image.
In the second step, the lateral expansion of hotspots along the i-coordinate is determined as the distance between the hotspot center and the first CCD pixel with $A_{i,j}<T_{hotspots}$. Since the spatial intensity distribution of individual hotspots can be described by a 2D Gaussian distribution, the expansion along the i-coordinate is sufficient.
The CCD pixel amplitudes $A_{i,j}$ attributed to one hotspot are summed up and interpreted as the light emission intensity of the hotspot. We call this quantity $I_{hotspots}^{single}$.
In the last step, the detected hotspots are ordered in descending order of $I_{hotspots}^{single}$. The number $N_{hotspots}$ is defined by $90\text{ \%}$ of the brightest hotspots. The additional cut is implemented in order to suppress false positives with small CCD-pixel amplitudes.
\\
The presented method is applicable under the condition of well separated hotspots. The emitted light intensity increases with $\Delta V$ as a consequence of the increasing $DCR$ and the increasing gain of the SiPM (see equation \ref{Eq_Emission}). 
The FWHM (full width at half maximum) of the lateral hotspot expansion does not show a significant increase with $\Delta V$ \cite{Popova2013}.
However, an absolute increase of the lateral expansion of hotspots is observed with $\Delta V$, as displayed in figure \ref{fig_Ampl_Distr_Two_HS}. 
Two hotspots cannot be separated if $A_{i,j}$ does not drop below $T_{hotspots}$ in between them. But since the observed $N_{hotsopts}$ is much lower than the number of SiPM micro-cells, we do not consider this a serious problem for our method. Figure \ref{fig_Number_Hotspots(Th_hotspots)} shows $N_{hotspots}$ as a function of $T_{hotspots}$ for the PM3350T STD at $\Delta V=5.4\text{ V}$. The vertical line indicates $T_{hotspots}$ that is chosen in this work.
\FloatBarrier
\subsection{Correlation between the dark count rate and the emitted light intensity}
\label{sub_Correlation_DCR_Light_Intensity}
\begin{figure}
\centering
\resizebox{0.5\textwidth}{!}{%
  \includegraphics{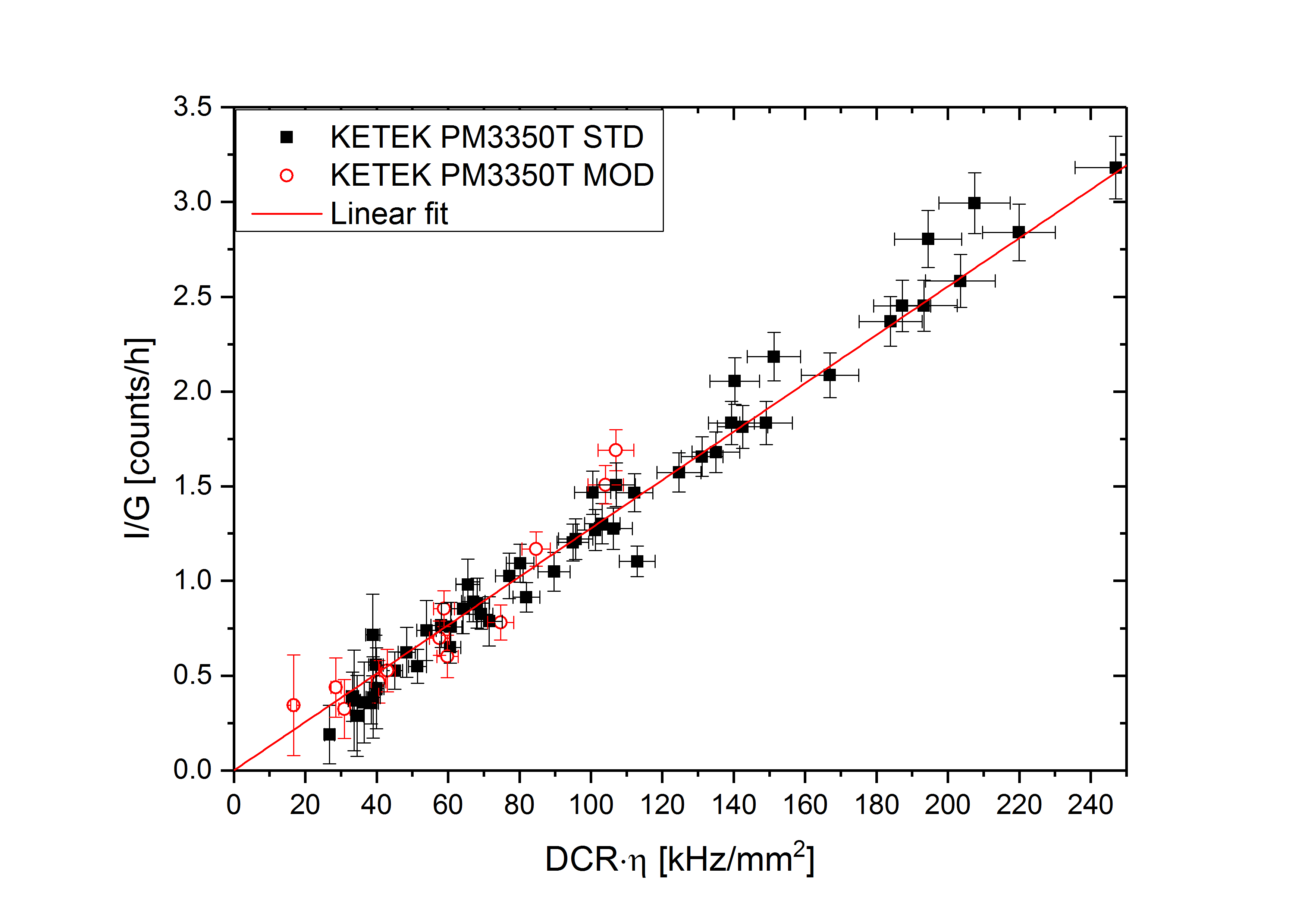}
}
\caption{Linear correlation between $(I/G)$ and $(DCR\cdot \eta)$.}
\label{I_tot_vs_DCR_CTP_corr}       
\end{figure}

\begin{figure}
\centering
\resizebox{0.5\textwidth}{!}{%
  \includegraphics{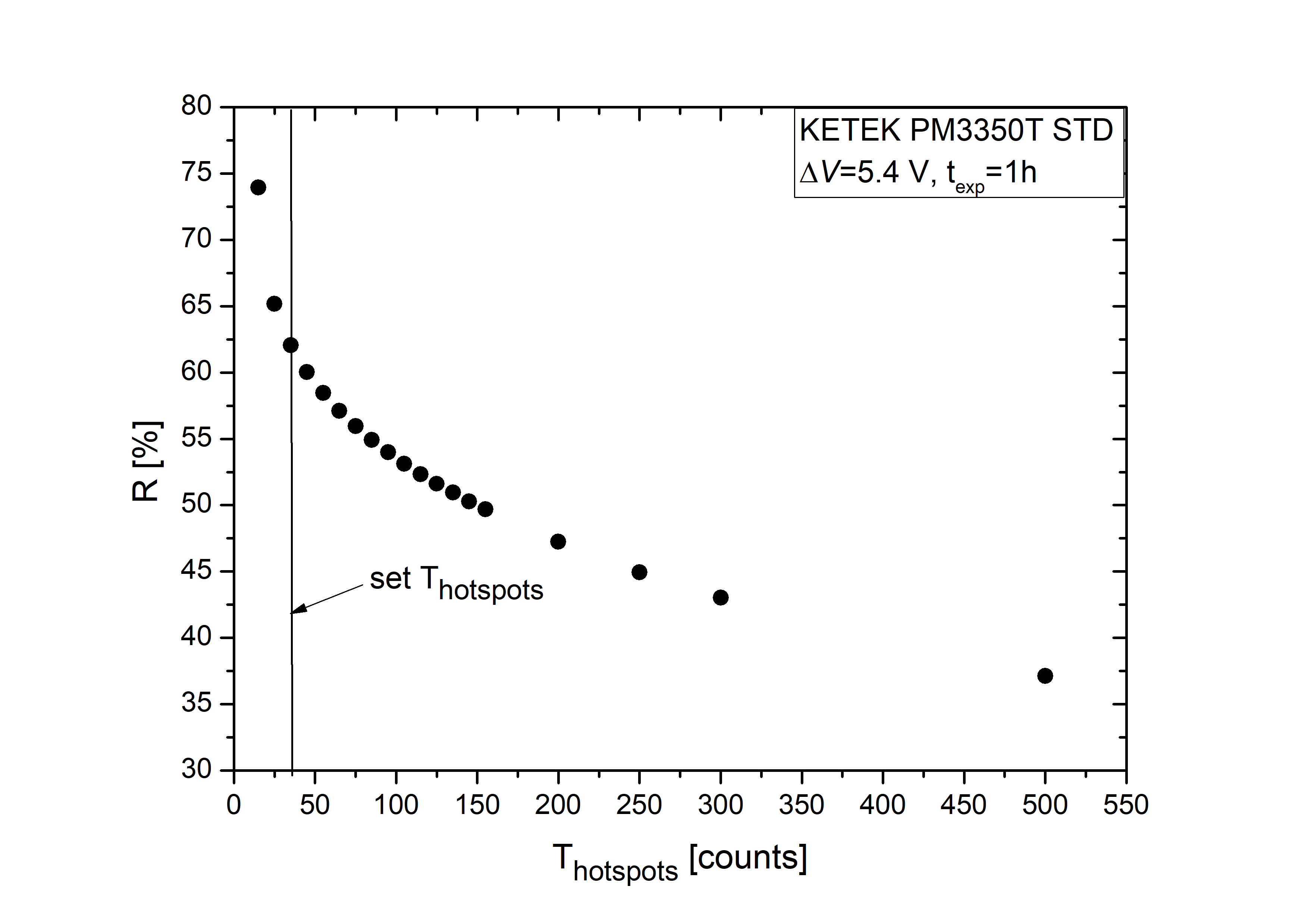}
}
\caption{Contribution $R$ of the hotspots to the total $DCR$ vs. $T_{hotspots}$.}
\label{fig_R(Th_hotspots)}       
\end{figure}

Equation \ref{Eq_Emission} gives an expression for the detected fraction $I$ of the total light intensity, which is emitted by a SiPM. Here, $t_{exp}$ is the exposure time, G is the gain of the SiPM, $DCR$ is the dark count rate, $\eta$ is the correction factor accounting for the effect of optical crosstalk, $PDE_{CCD}(\lambda)$ is the detection efficiency of the camera and $\kappa (\lambda)$ is the number of emitted photons per electron-hole pair during the avalanche breakdown of a micro-cell. 
\begin{equation}
\label{Eq_Emission}
I=t_{exp} \cdot G \cdot DCR \cdot \eta \cdot \int_{\lambda} \! PDE_{CCD}(\lambda)\cdot \kappa(\lambda) \, \mathrm{d}\lambda
\end{equation}
In order to measure the $DCR$, an analysis of digitized, randomly triggered, $1\text{ }\mu\text{s}$ long signal traces was performed.
$DCR$ is defined as the average number of pulses with an amplitude larger than $0.5\text{ photoelectrons (p.e.)}$ in one trace, divided by the length of the trace.
By the effect of optical crosstalk, secondary photons emitted during a primary dark pulse initiate additional avalanches in neighboring micro-cells. The additionally firing micro-cells then again produce secondary photons, which are detected by the CCD. In order to correlate $I$ and $DCR$, the number of crosstalk pulses following an initial primary dark pulse has to be taken into account.
\\
The optical crosstalk probability $P_{CT}$ is defined as the ratio of $DCR$ measured at an amplitude level of $1.5\text{ p.e.}$ and $0.5\text{ p.e.}$ \cite{Renker2009} (see equation \ref{eq_CTP_itimized}).
\begin{equation}
\label{eq_CTP_itimized}
P_{CT}=\frac{DCR(1.5\text{ p.e.})}{DCR(0.5\text{ p.e.})}
\end{equation}
In \cite{Vinogradov2012}, the optical crosstalk is analytically modeled as a branching Poisson process.
In that model every pulse produces a Poisson distributed number of crosstalk pulses with the mean $\lambda$. Equation \ref{eq_CTP_Mean_Number} gives the relation between $\lambda$ and $P_{CT}$. 
The total number of pulses originated from a single non-random primary event follows the Borel distribution with the mean $\eta$, which is given in equation \ref{eq_Total_Events_Mean_Number}. For the case of a Poisson distributed number of primary events, the pulse-height spectrum is described by a Generalized Poisson distribution.
\begin{equation}
\label{eq_CTP_Mean_Number}
\lambda=-ln\left(1-P_{CT}\right)
\end{equation}
\begin{equation}
\label{eq_Total_Events_Mean_Number}
\eta=\frac{1}{1-\lambda}
\end{equation}
The gain $G$ is measured by flashing the SiPM with a pulsed laser ($406\text{ nm}$, $1\text{ kHz}$ repetition rate, $<70\text{ ps}$ pulse width) which drives the SiPM into saturation (number of emitted photons $>>$ number of micro-cells).
For each flash the SiPM signal is integrated for a duration of $450\text{ ns}$ and the gain is determined by dividing the integral charge by the load resistance, the number of SiPM micro-cells, and the elementary charge \cite{Schneider}. The breakdown voltage $V_{bd}$ is obtained by extrapolating $G(V)$ to $G(V_{bd})=0$.
In figure \ref{I_tot_vs_DCR_CTP_corr}, both SiPMs show the same dependence of $(I/G)$ on
$(DCR\cdot \eta)$, which indicates the similarity of their emitted wavelength spectrum. The data points were obtained by varying $\Delta V$.
The observed linear correlation between the emitted light intensity and the dark count rate is in agreement with the observation reported in \cite{Lacaita1993} for an avalanche photodiode. This correlation allows us to generate a spatially resolved map of the dark count rate of a SiPM.
We use equation \ref{Ratio_R} to determine the contribution $R$ of hotspots to the measured $DCR$ of the SiPM. Figure \ref{fig_R(Th_hotspots)} shows how $R$ depends on the choice of $T_{hotspots}$ for the PM3350T STD at $\Delta V=5.4 \text{ V}$.
\begin{equation}
\label{Ratio_R}
DCR_{hotspots}=\frac{I_{hotspots}}{I_{hotspots}+I_{glowing}} \cdot DCR=R \cdot DCR
\end{equation}
\FloatBarrier
\section{Results}
\label{sec_Results}
\subsection{Temperature dependence of the emitted light intensity}
\label{sub_Temperature}
\begin{figure}
\centering
\resizebox{0.5\textwidth}{!}{%
  \includegraphics{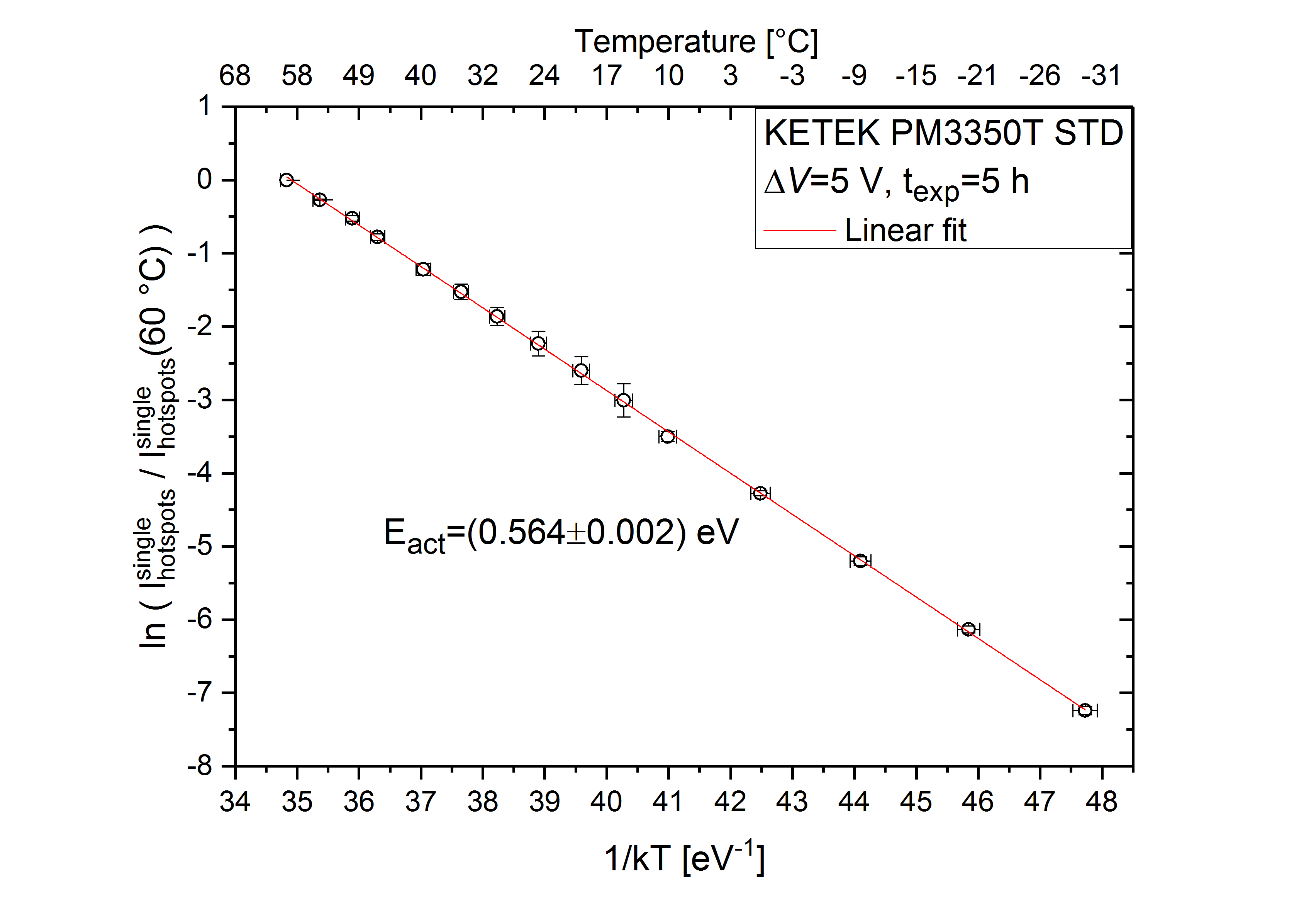}
}
\caption{Arrhenius plot of $I_{hotspots}^{single}(T)$ normalized to $I_{hotspots}^{single}(T=60\text{ }^{\circ}\text{C})$, for 5 random hotspots.}
\label{Arrhenius}       
\end{figure}

The dominating mechanism which is responsible for hot\-spots is identified by investigating the temperature dependence of $I_{hotspots}^{single}$.
The measurements are performed on the\\ 
PM3350T STD between $+60\text{ }^{\circ}\text{C}$ and $-30\text{ }^{\circ}\text{C}$. In order to adjust the temperature of the device, the SiPM is mounted onto a thermo-electric cooler inside an evacuated TO-8 housing.
The measurement is carried out at each temperature with an exposure time between $1\text{ h}$ and $5\text{ h}$, at $\Delta V=5 \text{ V}$.
For five randomly selected hotspot, the activation energy $E_{act}$ of the hotspot mechanism is determined by using the Arrhenius plot. $I_{hotspots}^{single}(T)$ of each hotspot is normalized to $I_{hotspots}^{single}(T=60\text{ }^{\circ}\text{C})$.
In figure \ref{Arrhenius} the Arrhenius plot of the normalized $I_{hotspots}^{single}(T)$ is shown. 
The slope of the linear fit yields an $E_{act}$ of $0.56\text{ eV}$, which is half the bandgap energy in silicon. The DCR generation mechanism can thus be attributed to the Shockley-Read-Hall-Generation \cite{Sze}.
We conclude from this result that the occurrence of hotspots in the tested SiPMs is caused by a spatially confined increase of the generation-recombination center density or the existence of crystal defect types with an enhanced charge-carrier generation rate. 
It is assumed that the ion-implantation process is responsible for the majority of crystal defects, which are introduced during the fabrication process. For this reason we single out the ion implantation as a process, which needs to be optimized in order to achieve a lower hotspot density and consequently a lower $DCR$.
\FloatBarrier
\subsection{Number of hotspots}
\label{sub_Number_of_hotspots}
\begin{figure}
\centering
\resizebox{0.5\textwidth}{!}{%
  \includegraphics{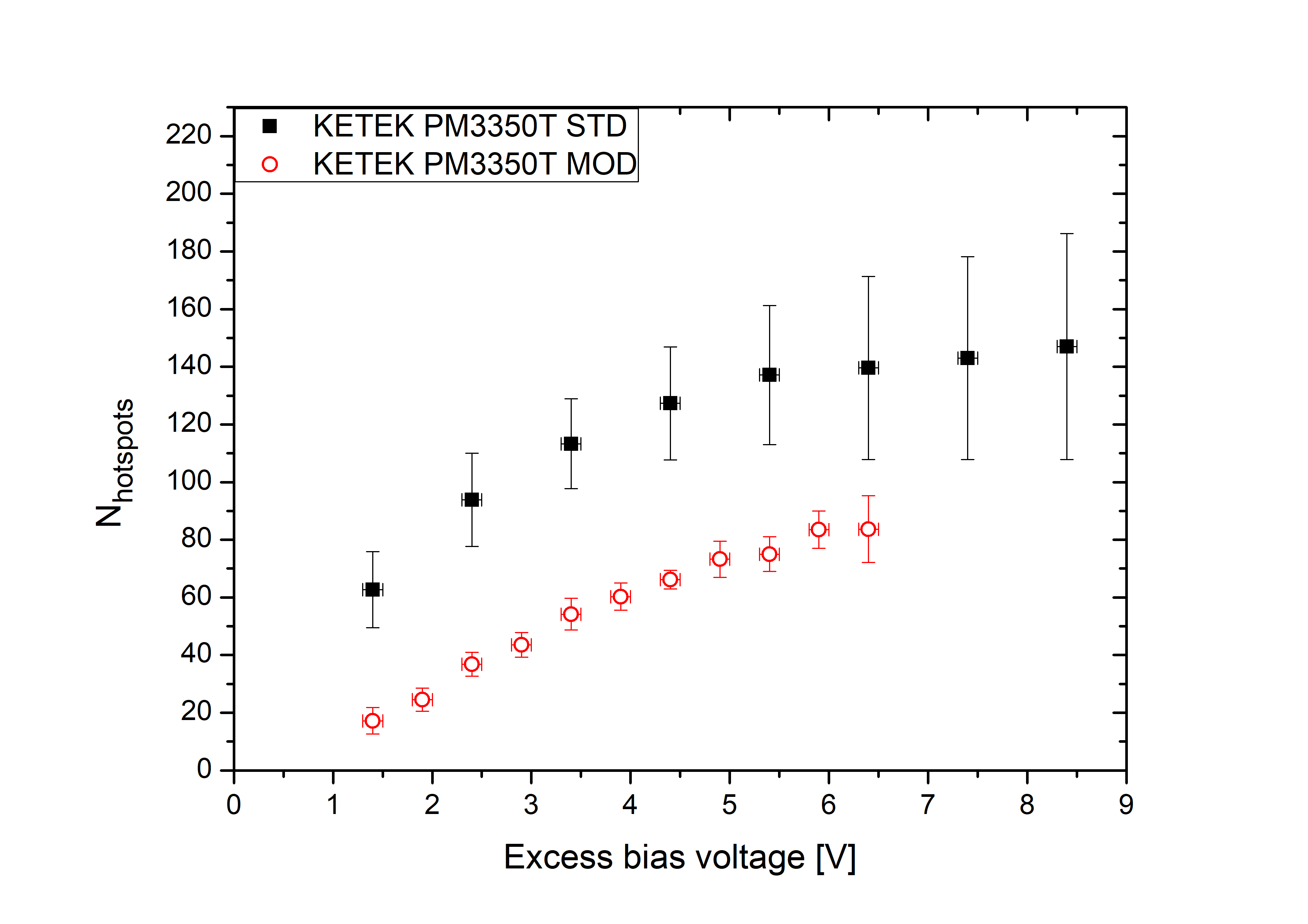}
}
\caption{Number of hotspots as a function of $\Delta V$.}
\label{Number_hotspots_all}       
\end{figure}

Figure \ref{Number_hotspots_all} shows $N_{hotspots}$ vs. $\Delta V$ for the PM3350T STD and the PM3350T MOD. Here, the total SiPM area is in the field of view of the CCD. When operated at $\Delta V=5.4\text{ V}$, the PM3350T MOD has $1.83$ times fewer $N_{hotspots}$ than the PM3350T STD. The increase of $N_{hotspots}$ with $\Delta V$ can be attributed to the increase of the Geiger-discharge probability $P_{G}$, which is proportional to the $PDE$ of the SiPM. 
$N_{hotspots}$ is expected to saturate when the position dependent $P_{G}$ saturates at high enough overvoltages. This condition could not be reached for every investigated device, since contributions as insufficient quenching, high-field effects, etc. start to dominate the performance of the SiPMs at high $\Delta V$. In this regime a rapid increase of the dark current is observed (see figure \ref{fig_IV_STD_HE}).
Since the depletion volume does not significantly increase with $\Delta V$, the total number of contributing defects is assumed to be constant.
\FloatBarrier
\subsection{Contribution of hotspots to the dark count rate}
\label{sub_Contribution_of_hotspots}
\begin{figure}[h!]
\centering
\resizebox{0.5\textwidth}{!}{%
  \includegraphics{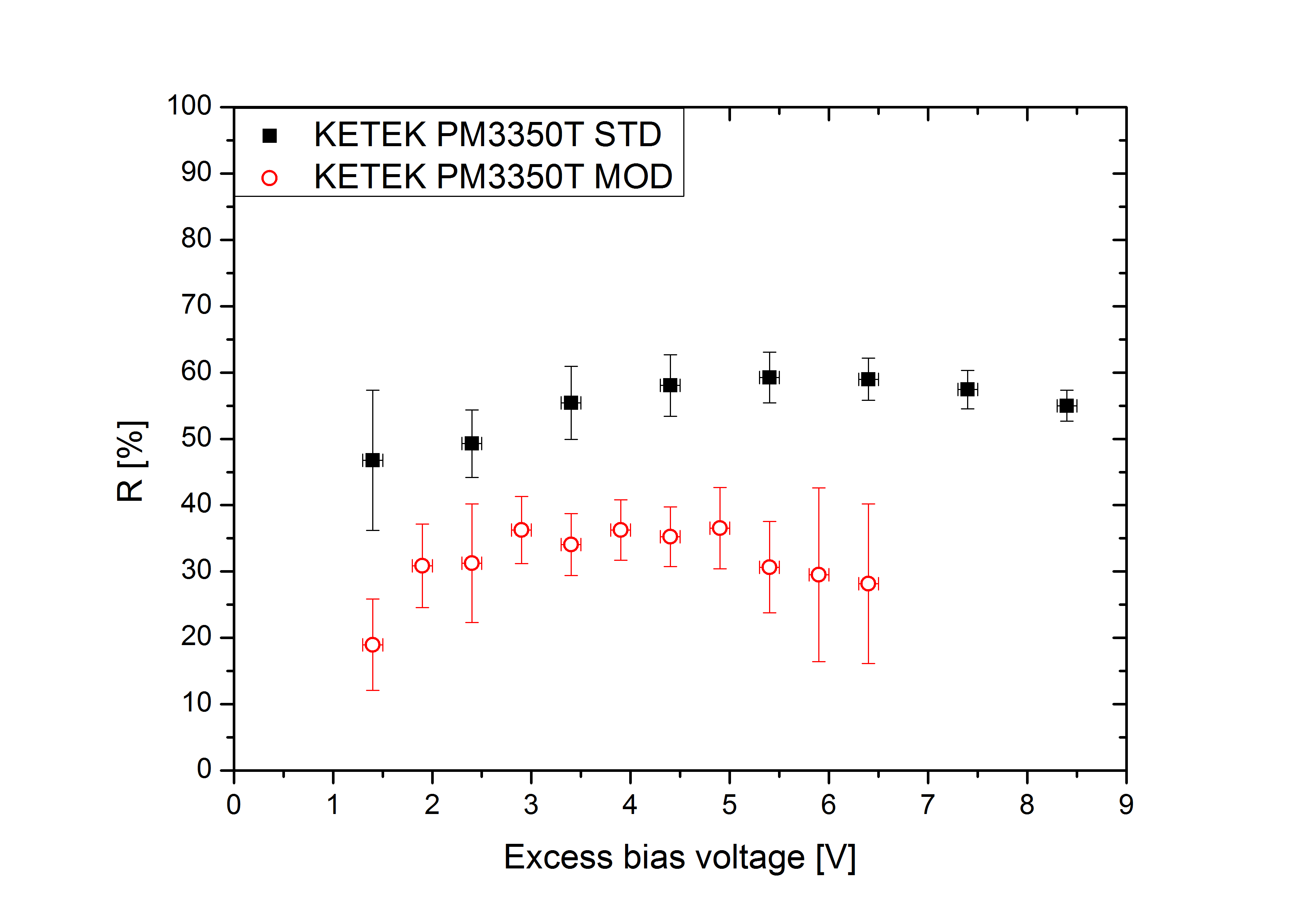}
}
\caption{Contribution $R$ of hotspots to the overall $DCR$ as a function of $\Delta V$.}
\label{Mean_R}       
\end{figure}

\begin{figure}
\centering
\resizebox{0.5\textwidth}{!}{%
  \includegraphics{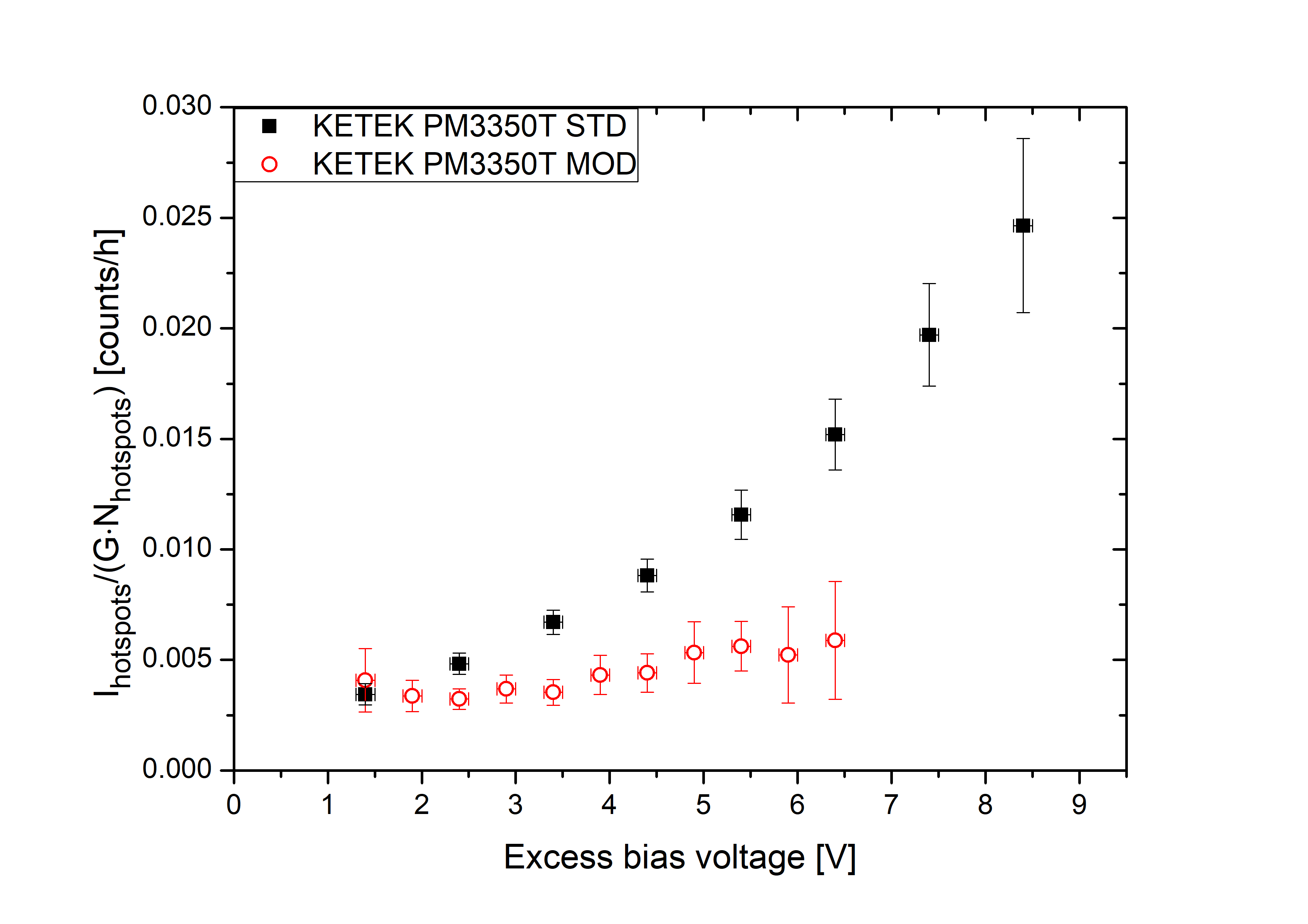}
}
\caption{Average light emission intensity per hotspot as a function of $\Delta V$.}
\label{Average_I_hotspots}       
\end{figure}

\begin{figure}
\centering
\resizebox{0.5\textwidth}{!}{%
  \includegraphics{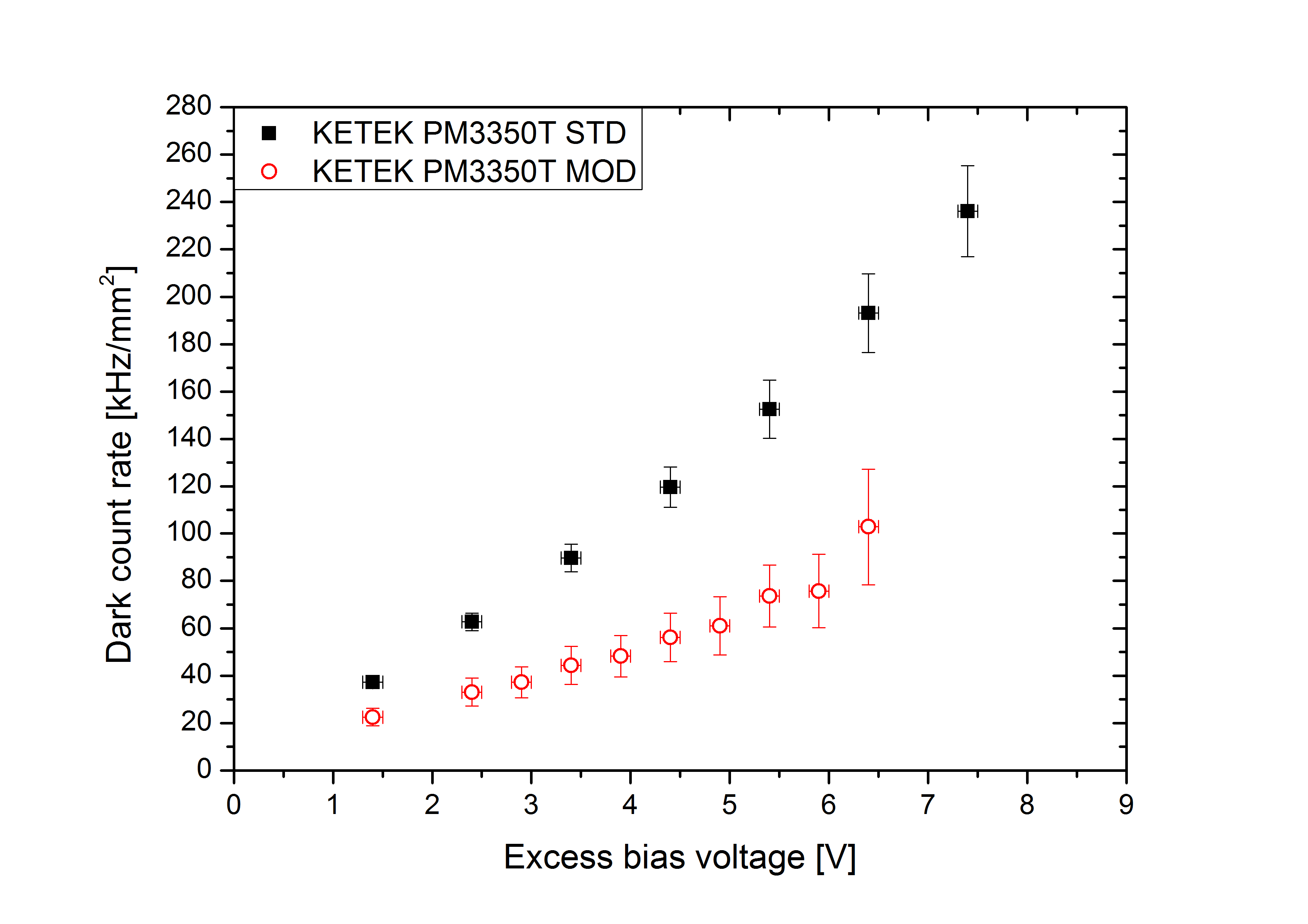}
}
\caption{Dark count rate as a function of $\Delta V$ at $(21\pm1)\text{ }^{\circ}\text{C}$.}
\label{DCR_vs_OV_STD_OPT1_OPT2}       
\end{figure}

\begin{figure}
\centering
\resizebox{0.5\textwidth}{!}{%
  \includegraphics{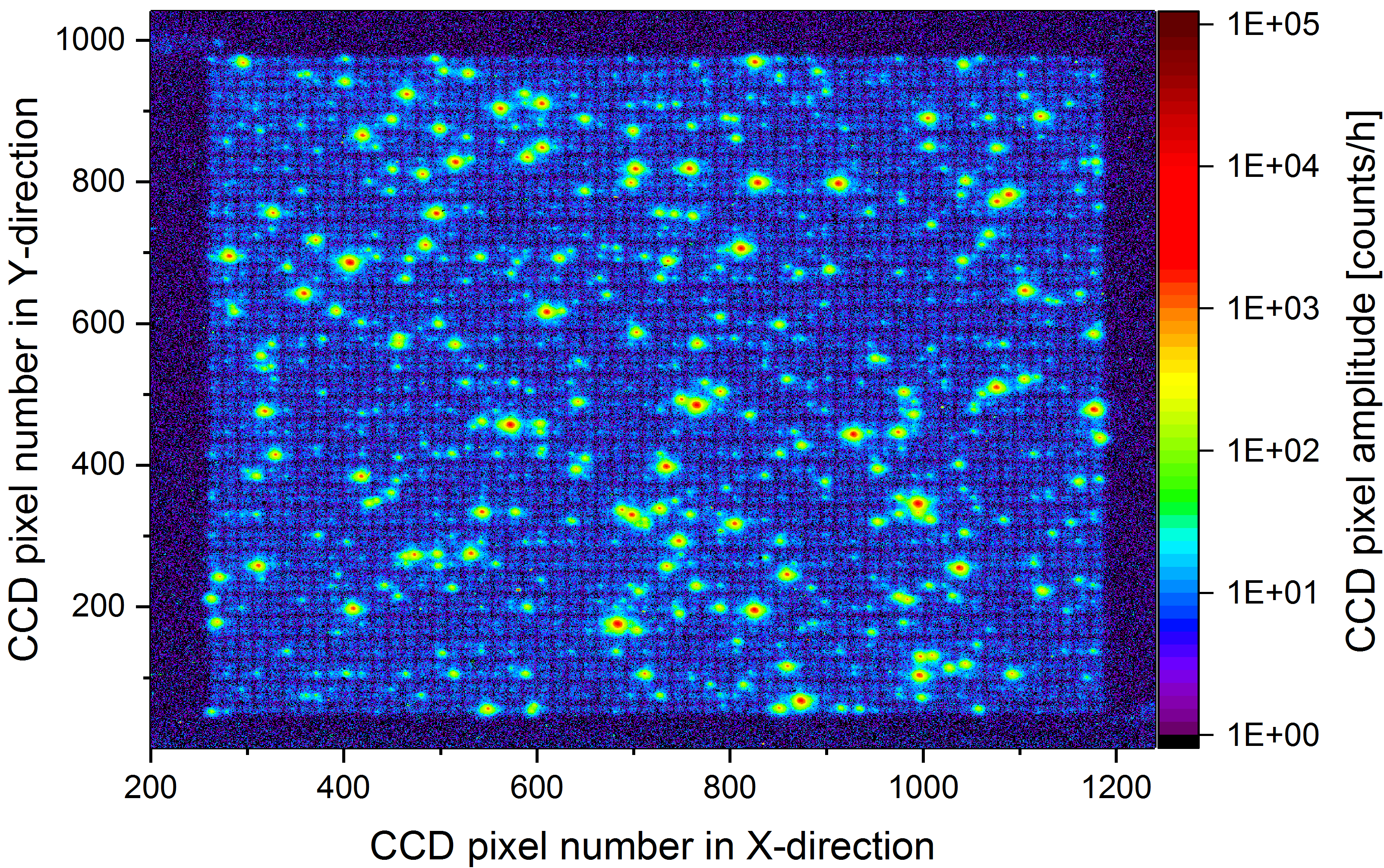}
}
\caption{Light emission image of the PM3350T STD at $\Delta V=5.4 \text{ V}$ and $t_{exp}=1\text{ h}$.}
\label{Clara_Pictures_a}       
\end{figure}

\begin{figure}
\centering
\resizebox{0.5\textwidth}{!}{%
  \includegraphics{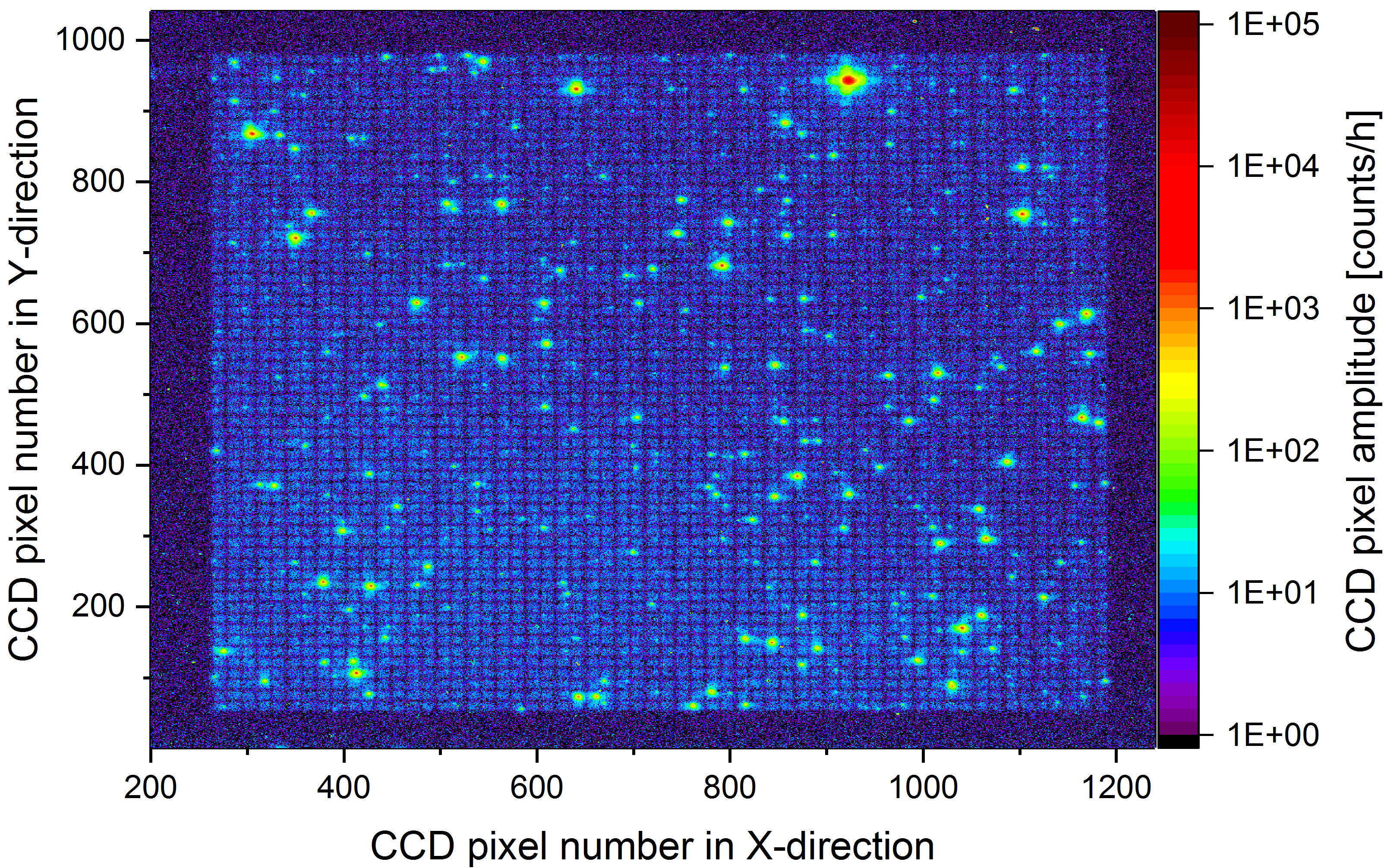}
}
\caption{Light emission image of the PM3350T MOD at $\Delta V=5.4 \text{ V}$ and $t_{exp}=1\text{ h}$.}
\label{Clara_Pictures_b}       
\end{figure}

In figure \ref{Mean_R} the contribution $R$ of hotspots to the total dark count rate of the tested SiPMs is shown. $R$ is observed to be independent of $\Delta V$ within the uncertainties. The determined contribution is reduced from about $56\text{ \%}$ for the PM3350T STD to about $33\text{ \%}$ for the PM3350T MOD.
\FloatBarrier
In order to compare the average light intensities emitted by a hotspot, $I_{hotspots}$ is normalized to the product of $G$, $N_{hotspots}$ and $t_{exp}$. The resulting quantity is proportional to the amount of photoelectrons produced in the CCD per electron-hole pair in the avalanche region, per hour and per hotspot. 
The normalized $I_{hotspots}$ for the PM3350T MOD is approximately two times lower with respect to the PM3350T STD, when operated at $\Delta V=5.4\text{ V}$ (see figure \ref{Average_I_hotspots}). This result leads us to the conclusion that in addition to the reduced number of contributing crystal defects, the dark count rate generated by these defects is also suppressed.
Due to the significant reduction of $R$, the $DCR$ at $\Delta V=5.4 \text{ V}$ is suppressed from $(152\pm 12)\text{ kHz/mm}^2$ for the PM3350T STD to $(73\pm 13)\text{ kHz/mm}^2$ for the PM3350T MOD at $21\text{ }^{\circ}\text{C}$ (see figure \ref{DCR_vs_OV_STD_OPT1_OPT2}).
Figures \ref{Clara_Pictures_a} and \ref{Clara_Pictures_b} show examples of light emission images at $\Delta V=5.4 \text{ V}$ for the PM3350T STD and the PM3350T MOD. The reduction in the number and the light emission intensity of hotspots of the PM3350T MOD is evident in these images. The contribution from $I_{glowing}$ is equal for both SiPMs when normalized to the gain.
\FloatBarrier
\section{Conclusion}
\label{sec_Conclusion}
In this work, we presented a method for a spatially resolved determination of the dark count rate of SiPMs with a sub-micro-cell resolution.
It is based on the detection of photons which are emitted by hot carriers during the Geiger-discharges of micro-cells.
In this article, we showed how the emitted light intensity is correlated with the dark count rate of the SiPM.
\\
In our work, we investigated two KETEK SiPM types with significantly different dark count rates (6 samples for each type).
For both SiPM types, we identified sub-micro-cell regions with an enhanced light emission intensity (hotspots).
The significant difference in the dark count rate of the investigated SiPMs was attributed to the difference in the hotspot density and the dark count rate per hotspot.
The dark count rate of micro-cell areas without hotspots was observed to be equal for both SiPM types.
\\
The temperature dependence of the emitted light intensity identifies the Shockley-Read-Hall-Generation as the dominating mechanism in hotspot-areas.
Consequently, we attribute the occurrence of hotspots to a spatially confined increase of the charge carrier generation rate.
\section{Acknowledgements}
This work was partially supported by the Russian grants 3.2989.2017/ 4.6 and 3.8484.2017 / 9.10. from the part of MEPHI.\\
We thank KETEK GmbH for providing the SiPM samples and the measurement equipment. 
We would like to express special appreciations to Dr. Florian Wiest and Dr. Peter Iskra from KETEK for fruitful discussions
and the support of the presented investigations.
%
 \bibliographystyle{elsarticle-num}
 \bibliography{mybibfile}

\begin{thebibliography}{10}
\expandafter\ifx\csname url\endcsname\relax
  \def\url#1{\texttt{#1}}\fi
\expandafter\ifx\csname urlprefix\endcsname\relax\def\urlprefix{URL }\fi
\expandafter\ifx\csname href\endcsname\relax
  \def\href#1#2{#2} \def\path#1{#1}\fi

\bibitem{Gautam}
{D. K. Gautam}, {K. B. Garg}, {W. S. Khokle}, {Photon Emission from
  Reverse-Biased Silicon P-N Junctions}, Solid-State Electronics 31 (1988)
  219--222.
\newblock \href {http://dx.doi.org/10.1016/0038-1101(88)90130-X}
  {\path{doi:10.1016/0038-1101(88)90130-X}}.

\bibitem{Lacaita1993}
A.~L. Lacaita, F.~Zappa, S.~Bigliardi, M.~Manfredi, {On the Bremsstrahlung
  Origin of Hot-Carrier-Induced Photons in Silicon Devices}, IEEE Transactions
  on Electron Devices 40~(3) (1993) 577--582.
\newblock \href {http://dx.doi.org/10.1109/16.199363}
  {\path{doi:10.1109/16.199363}}.

\bibitem{Mirzoyan}
{R. Mirzoyan}, {R. Kosyra}, {H.G. Moser}, {Light emission in Si avalanches},
  Nuclear Instruments and Methods in Physics Research A 610 (2009) 98--100.
\newblock \href {http://dx.doi.org/10.1016/j.nima.2009.05.081}
  {\path{doi:10.1016/j.nima.2009.05.081}}.

\bibitem{dSiPM}
{ T. Frach}, {G. Prescher}, {C. Degenhardt}, {R. de Gruyter}, {A. Schmitz}, {R.
  Ballizany}, {The Digital Silicon Photomultiplier - Principle of Operation and
  Intrinsic Detector Performance}, IEEE Nuclear Science Symposium Conference
  Record N28-5 (2009) 1959--1965.
\newblock \href {http://dx.doi.org/10.1109/NSSMIC.2009.5402143}
  {\path{doi:10.1109/NSSMIC.2009.5402143}}.

\bibitem{KETEK}
{KETEK GmbH}, webpage: \url {https://www.ketek.net/} (accessed May 02, 2018).

\bibitem{Andor}
{Andor Clara, Datasheet}, online available at: \url
  {http://www.andor.com/pdfs/specifications/Andor_Clara_Series_Specifications.pdf}
  (accessed May 02, 2018).

\bibitem{Eckert2010}
P.~Eckert, H.~Schultz-Coulon, {Characterisation Studies of Silicon
  Photomultipliers}, {Nuclear Instruments and Methods in Physics Research A}
  620 (2010) 217--226.
\newblock \href {http://dx.doi.org/10.1016/j.nima.2010.03.169}
  {\path{doi:10.1016/j.nima.2010.03.169}}.

\bibitem{Otte2016}
A.~N. Otte, D.~Garcia, T.~Nguyen, D.~Purushotham, {Characterization of Three
  High Efficiency and Blue Sensitive Silicon Photomultipliers}, Nuclear
  instruments and methods in physics research A 846 (2016) 106--125.
\newblock \href {http://dx.doi.org/10.1016/j.nima.2016.09.053}
  {\path{doi:10.1016/j.nima.2016.09.053}}.

\bibitem{Popova2013}
{E. Popova et al.}, {Simulation and measurements of Geiger discharge transverse
  size in a SiPM cell}, in: Talk given at IEEE NSS/MIC/RTSD, Seoul, 2013.

\bibitem{Renker2009}
D.~Renker, E.~Lorenz, {Advances in solid state photon detectors}, Journal of
  Instrumentation 4 (2009) P04004.
\newblock \href {http://dx.doi.org/10.1088/1748-0221/4/04/P04004}
  {\path{doi:10.1088/1748-0221/4/04/P04004}}.

\bibitem{Vinogradov2012}
S.~Vinogradov, {Analytical models of probability distribution and excess noise
  factor of solid state photomultiplier signals with crosstalk}, Nuclear
  Instruments and Methods in Physics Research, Section A: Accelerators,
  Spectrometers, Detectors and Associated Equipment 695 (2012) 247--251.
\newblock \href {http://dx.doi.org/10.1016/j.nima.2011.11.086}
  {\path{doi:10.1016/j.nima.2011.11.086}}.

\bibitem{Schneider}
F.~R. Schneider, T.~R. Ganka, G.~Seker, E.~Engelmann, D.~Renker, S.~Paul,
  W.~Hansch, S.~I. Ziegler, {Characterization of blue sensitive 3x3 mm2 SiPMs
  and their use in PET}, Journal of Instrumentation 9~(07) (2014) P07027.
\newblock \href {http://dx.doi.org/10.1088/1748-0221/9/07/P07027}
  {\path{doi:10.1088/1748-0221/9/07/P07027}}.

\bibitem{Sze}
{S. M. Sze}, {K. K. NG}, {Physics of Semiconductor Devices}, John Wiley \& Sons
  Inc., 2007.

\end{thebibliography}
%
\end{document}